# An Assessment of the Space Radiation Environment in a Near-Equatorial Low Earth Orbit Based on the RazakSAT-1 Satellite

Wayan Suparta[*], Siti Katrina Zulkeple

Space Science Centre (ANGKASA), Institute of Climate Change,
Universiti Kebangsaan Malaysia, 43600, Bangi, Selangor Darul Ehsan, Malaysia

## Abstract

The Malaysian satellite RazakSAT-1 was designed to operate in a near-equatorial orbit (NEqO) and low earth orbit (LEO). However, after one year of operation in 2010, communication to the satellite was lost. This study attempted to identify whether space radiation sources could have caused the communication loss by comparing RazakSAT-1 with two functional satellites. Data on galactic cosmic rays (GCR), trapped protons, trapped electrons, and solar energetic particles (SEPs) obtained from Space Environment Information System (SPENVIS) was analyzed. The effects of these radiation sources were analyzed using three parameters: linear energy transfer (LET), total ionizing dose (TID) and solar cell degradation for a three-year mission (July 2009 – July 2012). We also considered the flux data from National Oceanic and Atmospheric Administration (NOAA) 15, 16, and 17 and the geomagnetic conditions during the time when RazakSAT-1's communication was likely lost (June to August 2010). Within the LEO/NEqO, GCR dominated at a high energy range of $10^3$ MeV and above, whereas the energies of the trapped protons and trapped electrons were less than 400 MeV and 4 MeV, respectively. Meanwhile, no SEPs were estimated. For the radiation doses, we found that Satélite de Coleta de Dados (SCD-2) and Advanced Land Observation Satellite (ALOS) were exposed to higher radiation damage than RazakSAT-1, yet they remained functional. The reasons for the communication disruption of RazakSAT-1 were further discussed.

[*] Corresponding author: Tel.: +603 8911 8485; Fax: +603 8911 8490
 E-mail address: wayan@ukm.edu.my (Wayan Suparta)

*Keywords*: Low earth orbit, linear energy transfer, total ionizing dose, non-ionizing energy loss

**1. Introduction**

RazakSAT-1, which was launched in 2009, is Malaysia's second remote sensing satellite [1]. The satellite was designed to operate in a low earth orbit (LEO) and near-equatorial orbit (NEqO) at an altitude of 685 km with a 9° inclination. This orbital position allows for a satellite revisit frequency of 90 minutes, which is more frequent than a sun-synchronous orbital satellite. At this altitude, RazakSAT-1 can monitor the equatorial environment.

The payload in RazakSAT-1 is a medium-sized aperture camera (MAC) with five linear detectors, each of which consists of one panchromatic band and four multispectral bands that are designed to capture high-resolution images [2]. The high-resolution images contribute to Malaysia's development by aiding land management, resource development and forestry. The satellite's high-speed transmission system can download 32 gigabits of stored image data within 3-4 daytime passes [3]. Power is supplied by solar cells made of GaAs/Ge, whereas the battery uses NiCd. Overall, RazakSAT-1 had a positive outlook, but communication to the satellite gradually began to fade out before it was completely lost a year later. The true cause of the disruption was unknown because the vehicle telemetry was unavailable. Our work was motivated by the unexpected failure of RazakSAT-1. Specifically, we aimed to determine whether the space radiation environment had a role in the communications failure.

[4] forecasted the 2012 space environment for the newly launched EGYPTSAT-2 satellite to assist the satellite team in proper planning decisions using the models in the Space Environment Information System (SPENVIS). The authors predicted that for a 1.5 mm aluminum shielding thickness, the five-year mission would provide a total ionizing dose (TID) of $2.65 \times 10^4$ rads through the silicon material and a displacement damage dose (DDD) of $7.75 \times 10^7$ MeV/g (Si). Because these values are less than the critical threshold, the satellite is not expected to experience critical damage. However, EGYPTSAT-2 is a LEO satellite that orbits the polar region.

Our work examined space radiation, i.e., galactic cosmic ray (GCR), trapped particles and solar energetic particles (SEPs), in the LEO/NEqO as a plausible cause of the failure of RazakSAT-1 using the SPENVIS online tool. The orbital data of RazakSAT-1 was compared with the higher inclination of the Satélite de Coleta de Dados (SCD-2) and the polar Advanced Land Observation Satellite (ALOS), which may operate well beyond their mission lifetimes. In addition, this work provides additional information on the natural characteristics of the LEO/NEqO environment.

This paper is divided into three sections. The first section presents the methodology, including the models that were used to estimate the fluxes of GCR, SEPs and trapped particles, the radiation doses, i.e., linear energy transfer (LET) and TID, and degradation of the solar cell. We also examined the geomagnetic conditions using National Oceanic and Atmospheric Administration / Polar Operating Environmental Satellites (NOAA/POES) data. The second section discusses the results, which we further divided into five subsections: analysis of fluxes, LET, TID, solar-cell degradation and geomagnetic conditions. Finally, the results are summarized, and the work is concluded in the third section.

## 2. Methodology

### 2.1 Estimating Fluxes

The models to predict the space environment of the LEO/NEqO are shown in Figure 1. These models are available in SPENVIS, which is an online tool that contains many integrated models designed for various simulation-based studies [5]. All of these models were run for magnetic storms to provide the worst-case scenario.

Based on Figure 1, we first estimated the fluxes of GCR, SEP, trapped protons and trapped electrons in the RazakSAT-1 satellite orbit. The RazakSAT-1 satellite orbit was set at 9° inclination with a perigee of 662 km and an apogee of 687 km for three-year mission from 14 July 2009 to 14 July 2012. We also compared RazakSAT-1 with two satellites: the SCD-2 satellite and ALOS, which were functional during the mission period. SCD-2 is a data-collection satellite from Brazil under the program of the Instituto Nacional de Pesquisas Espaciais (National Institute of Space Research or INPE), whereas ALOS is an earth observation satellite developed by the Japan Aerospace Exploration Agency (JAXA). Orbital data on these satellites are simplified in Table 1. Both RazakSAT-1 and SCD-2 satellites are located in a NEqO, whereas ALOS is located in a polar orbit. Therefore, different inclinations can be compared.

To generate GCR fluxes, we used the CREME96 model because it includes anomalous cosmic ray and low-energy components that are not available in other models. Anomalous cosmic rays (ACR) are partially ionized particles with low energy that can be trapped in the earth's inner radiation belt due to its single-charge nature. In addition, CREME96 utilizes the solar minimum data from 1977 to estimate the GCR fluxes. We also compared the four available GCR models in SPENVIS (ISO 15390, CREME86, CREME96 and Nymmik) to identify their abilities to generate GCR fluxes in NEqO and polar orbits.

The ESP-PSYCHIC [6] model was used to estimate long-term SEP flux fluence over the three years. This model is advantageous because its data were taken from various satellites in three solar cycles, which allowed the model to predict a wide range of energy, from 1 MeV to above 100 MeV. The SEPs were estimated during a solar maximum and a magnetic storm.

Finally, NASA models AP-8 and AE-8 were selected to generate fluxes of trapped protons and trapped electrons, respectively. These models are standard models for trapped particles because they cover the entire energy range of trapped particles. AP-8 can cover an energy range of 0.1 to 400 MeV for trapped protons [7], whereas AE-8 can cover an energy range of 0.04 to 7 MeV for trapped electrons [8]. In this study, the trapped particles were estimated during a solar minimum. Then, the fluxes were applied to calculate the radiation doses.

## 2.2  Radiation Dose Calculation

The effects of space radiation are described as single-event effects (SEEs) and cumulative effects. SEE used LET parameters, whereas cumulative effects used TID parameters. We also included solar-cell degradation, which involved a type of cumulative effect called non-ionizing energy loss (NIEL).

### 2.2.1  Linear Energy Transfer

The LET spectrum inside the spacecraft is derived from differential energies that are generated from the models. The differential energy *f(E)* inside the spacecraft is converted into the differential LET *f(S)* using the CRÈME method [9]:

$$f(S) = f(E) \, dE/dS \qquad (1)$$

The differential energy *f(E)* is defined as:

$$f(E) = f'(E') \left[\frac{S(E')}{S(E)}\right]^{-\sigma} \qquad (2)$$

where *f'(E')* is the differential energy at the outer shell of the spacecraft, *S(E')* and *S(E)* are the stopping powers of the ions with energies *E'* and *E*, respectively, and *t* is the thickness of the spacecraft in g/cm² of Al or equivalent Al, and is defined as:

$$\sigma = \frac{\left[5 \times 10^{-2} \, A_g \left(A_n^{1/3} + 27^{1/3} - 0.4\right)^2\right]}{27} \qquad (3)$$

where $A_g$ is Avogadro's number and $A_n$ is the atomic mass of the incident ion.

Then, the LET spectrum is used to determine the single-event upset (SEU), which is a soft error that commonly occurs in the electronics of orbiting spacecraft. However, because different energies can produce similar LETs, the slowing- and stopping-ion algorithms are fed into the calculation to determine the upset rate. The upset rate *U*, which is calculated from the algorithm, is defined as:

$$U = \pi \sum_{z=1}^{9} \left[\int_{p_m}^{p_m} D(p) \int_{E_0}^{E_m} F(E,Z) d \right] \qquad (4)$$

where *A* is the surface area of the sensitive volume, *D(p)* is the differential path-length distribution in the sensitive volume (cm²g⁻¹), and *F(E,Z)* is the differential ion energy flux (m⁻²sr⁻¹ s⁻¹ MeV⁻¹). $P_{min}$ is the minimum path length for depositing the minimum energy to cause an upset. As such, only energy above the minimum energy can cause an upset. Meanwhile, $E_{max}$ corresponds to the maximum energy in the ion spectrum.

By applying the above equations through SPENVIS, the LET and SEU rates were determined in a gallium arsenide (GaAs) device, which was shielded with 5 mm Al. Then, the LET and SEU rates in the orbits of RazakSAT-1, SCD-2 and ALOS were compared. The source of the stopping power, which was included in the calculations, was obtained from [10].

### 2.2.2 Total Ionizing Dose

In SPENVIS, a computer code called SHIELDOSE is used to determine the radiation dose as a function of depth in a shielded material [11]. However, in this work, we used the upgraded version called SHIELDOSE-2 to calculate the TID because it better represents proton nuclear interactions and includes more choices of target materials [12]. The dose was measured from a detector at the center of an Al sphere by varying the thickness. The target materials were GaAs and silicon (Si).

When incident particles hit geometric shapes, they are calculated differently. Electrons were calculated with the Monte Carlo code ETRAN, which also considered the Bremsstrahlung effect. A straight-ahead, continuous-slow-down approximation was applied to protons using the stopping power and range data of [13]. However, the proton calculation did not include angular deflections and range straggling, which is negligible in spare-shielding calculations [14]. The doses were estimated for different orbital data of RazakSAT-1, SCD-2 and ALOS over three years. The results were subsequently compared.

### 2.2.3 Solar-Cell Degradation

In SPENVIS, solar-cell degradation can be estimated using two application programs: EQFLUX [15] and MC-SCREAM [16]. EQFLUX uses the damage-equivalent approach with a 1 MeV electron fluence to estimate the solar-cell degradation, i.e., how much estimated fluence in the selected orbit can cause damage similar to a 1 MeV electron fluence. MC-SCREAM uses the DDD method, which applies the NIEL parameter to estimate the end-of-life (EOD) solar-cell performance. For both programs, we estimated the degradation of the GaAs material that was shielded with various cover-glass thicknesses. GaAs was selected because it was the material of the solar cell in RazakSAT-1. Then, we compared the solar-cell degradation in three orbital datasets of RazakSAT-1, SCD-2 and ALOS for the three-year period.

### 2.2.4 Geomagnetic Conditions

Approximately one year after its launch in July 2009, RazakSAT-1's communication was lost. The study period was set to three months (June to August 2010). To evaluate the geomagnetic conditions during this approximate period, we compared the trapped-particle fluxes that were obtained from the Medium Energy Proton and Electron Detector (MEPED) sensors onboard three NOAA/POES satellites (NOAA-15, 16, and 17) with geomagnetic indices of $K_p$ and $Dst$ (disturbance storm time). The data were obtained from the NOAA website at http://www.ngdc.noaa.gov/stp/satellite/poes/dataaccess.html. Both the $K_p$ and $Dst$ indices were obtained from the International Service of Geomagnetic Indices (ISGI). The $K_p$ index was derived over a three-hour range, which was standardized from 13 observatories, whereas the $Dst$ index was derived hourly from four magnetic observatories.

## 3. Results and Discussion

### 3.1 Flux Analysis

GCR, SEP and trapped particles were determined using their respective models, as shown in Figure 1. For cosmic rays, four models that generate GCR fluxes are available in SPENVIS. The fluxes over the three-year mission are shown in Figure 2 for the NEqO (Figure 2a) and polar orbit (Figure 2b). The NEqO was represented by RazakSAT-1 orbital data, whereas the polar orbit was represented by ALOS orbital data. In the NEqO, all four models produced similar GCR fluxes despite using different datasets. Here, CREME96 generated the highest fluxes, whereas the fluxes of CREME86 were slightly lower. The lowest values were generated by ISO 15390 and Nymmik; these two models generated identical fluxes throughout the entire energy range and fluxes were similar to other models at $10^4$ MeV and above. In the comparison of fluxes between NEqO and polar orbits, we observed that the model selection was more crucial for the polar orbit than for the NEqO. In contrast to the NEqO, the models showed significant discrepancies for GCR in the polar orbit.

As shown in Figure 2b, all four models were consistent at energies above 10 MeV, but they provided different fluxes below 10 MeV. At this point, ISO 15390 produced the lowest fluxes among the models, which resulted in a nominal graph. CREME86 produced constant flux values for energies below 10 MeV. Both the CREME96 and Nymmik models had identical graph patterns, in which Nymmik produced the highest fluxes. However, because Nymmik did not account for ACR [17], the CREME96 model was selected to generate the GCR fluxes.

Using the CREME96 model, the estimated polar GCR (Figure 2b) had a wide range of energies of 1 MeV to $2 \times 10^4$ MeV, whereas NEqO GCR (Figure 2a) only has energy above $5 \times 10^3$ MeV. In other words, only protons with extremely high energies can penetrate the earth's magnetic field to reach low altitudes near the equator because no low-energy GCR was estimated in the NEqO. The low-energy particles deflected to follow the earth's magnetic field lines, converging at the poles; this phenomenon explains why the polar region has a wide range of GCR energies. Moreover, the GCR flux in the NEqO for the entire three-year mission did not exceed 0.3 $m^2 sr^{-1} s^{-1} (MeV/n)^{-1}$, whereas the polar orbit contained GCR values up to $10^3$ $m^2 sr^{-1} s^{-1} (MeV/n)^{-1}$, which is approximately 3,333 times greater than the fluxes in the NEqO.

The GCR fluxes that were estimated for the three satellites (i.e., different orbits) are compared in Figure 3. Here, RazakSAT-1 received the lowest flux because of its low inclination, followed by the SCD-2 satellite and ALOS, which received the highest flux. The maximum fluxes that RazakSAT-1, SCD-2 and ALOS received over the entire three-year mission were 2845.48 $m^2sr^{-1}\ s^{-1}$, 4845.96 $m^2sr^{-1}\ s^{-1}$ and 45622.80 $m^2sr^{-1}\ s^{-1}$, respectively. Hence, the GCR fluxes in the orbits of SCD-2 and ALOS were 1.70 and 16.03 times higher, respectively, than that of RazakSAT-1. The model indicates that SCD-2 and ALOS are more exposed to GCR particles than RazakSAT-1 because of their higher inclinations.

On the other hand, SEPs were not estimated for the NEqO but were estimated for the polar orbit using ALOS orbital data as shown in Figure 4. Values were not obtained for the particles in the RazakSAT-1 and SCD-2 satellite orbits from any of the models in SPENVIS; therefore, this indicates that the particles are deflected by the geomagnetic cutoff near the equator such that none reach the lower altitudes. Thus, SEPs have a higher chance of reaching the earth's poles than the equator when large fluences are involved [18]. Low-energy SEPs are more common than high-energy SEPs, and the estimated maximum flux is approximately $2.5 \times 10^{13}$ $cm^{-2}\ MeV^{-1}$ for a one-year period.

In terms of trapped particles, Figure 5 shows the energy spectra for trapped protons (Figure 5a) and trapped electrons (Figure 5b) in the RazakSAT-1 orbit. The energy of trapped protons (Figure 5a) is several MeV to several hundred MeV because of its dominance in the inner belt. However, the energy of trapped electrons in the LEO/NEqO is 100 times smaller than the energy range of trapped protons, as shown in Figure 5b. The differential fluxes for trapped protons over the three-year mission may increase to 230 $cm^{-2}\ s^{-1}(MeV)^{-1}$, whereas the differential fluxes for the trapped electrons may increase to 25,000 $cm^{-2}s^{-1}(MeV)^{-1}$. However, in terms of orbital inclinations, Figure 6 which compares the trapped particles among the three satellites, shows that RazakSAT-1 had the smallest estimated fluxes of trapped particles. The fluxes of trapped protons in RazakSAT-1 (Figure 6a) can differ by approximately 8 to 10 times with SCD-2 and 100 times with ALOS. Furthermore, for trapped electrons (Figure 6b), the flux in RazakSAT-1 can differ up to 650 and 1,000 times from those of SCD-2 and ALOS, respectively. Moreover, the maximum estimated energy range of SPENVIS models for trapped electrons in NEqO is 4 MeV, whereas for polar orbits, their energy can increase to 6 MeV. All four figures (Figures 3-6) show a particle decrease at higher energy levels but the level of radiation damage produced by all of these particles can only be explained using the parameters that are described in the following sections.

### 3.2 LET Analysis

The effect of space particles on materials in the spacecraft was estimated using radiation dose parameters. Figure 7 shows the LET spectra through a GaAs device that was shielded with 5 mm Al according to three orbital datasets; the spectra decrease as LET increases. The decrease in the LET spectra indicates that the particles' energies mostly exhibited a lower LET range. Over three years, RazakSAT-1 was estimated to have the lowest LET damage (the maximum LET was 1,087 MeV cm$^2$ g$^{-1}$), whereas SCD-2 and ALOS had maximum LETs of 1,162 MeV cm$^2$ g$^{-1}$ and 21,277 MeV cm$^2$ g$^{-1}$, respectively. For a 5 mm shielding, the estimated overall LET fluences of RazakSAT-1 were 1.3-9.4 times less than those of SCD-2 and approximately 3-120 times less than those of ALOS.

Although LET depends on the particle type and energy, it is used to predict the effect of a single event as a result of particle interactions with the target material. The effects include direct ionization and proton-induced ionization. The former is usually caused by heavy ions, whereas the latter involves the formation of secondary particles. Through LET, the protons directly interact with the target materials to cause ionization. Proton-induced ionization or nuclear interactions occur when the protons transfer energy to the target nuclei, which results in recoil particles that contribute to further ionization. These two mechanisms lead to SEU events. The predicted SEU rates in the GaAs device that was shielded with 5 mm

Al over three years are shown in Table 2. RazakSAT-1 was estimated to have the lowest probability of SEU events compared with the other satellites. In some cases, increasing the shielding results in the formation of secondary particles that will add to the dose; if so, then the nuclear interactions will significantly multiply these secondary particles. This scenario appears to occur for all three satellites. Therefore, ionization by nuclear interactions is greater than direct ionization; thus, protons are the major contributor to the ionization effect in the GaAs device, as opposed to heavy ions. By comparing the three satellites, the estimated total SEU rates of RazakSAT-1 are 8.01 and 5.33 times smaller than those of SCD-2 and ALOS, respectively. A 5 mm Al shielding can reduce the SEU rates by 2.70%, 2.48% and 19.35% for RazakSAT-1, SCD-2 and ALOS, respectively. However, among the three satellites, SCD-2 demonstrates the highest SEU rates because of its frequent visits over the South Atlantic Anomaly (SAA), where proton exposure is higher.

### 3.3 TID Analysis

Figure 8 shows the dose-depth curve of the GaAs material as a result of interacting with trapped protons of LEO/NEqO RazakSAT-1 during a solar minimum. Based on the figure, three types of radiation were estimated for the total dose: electrons, Bremsstrahlung radiation and trapped protons. The dose decreases with increasing shielding thickness; however, for trapped protons, increasing the shielding by approximately 4 mm or more is not significant. Adding more shielding is less effective and is possibly inapplicable in space designs because of high construction costs and mass/volume issues [19]. At approximately 6 mm shielding and above, electrons do not contribute to Bremsstrahlung particles as their curve decreases. However, at this point, the effect of Bremsstrahlung particles is not severe.

Figure 9 shows the dose comparison among the three satellites. RazakSAT-1 had the lowest estimated TID, whereas ALOS had the highest dose and SCD-2 had a moderate dose.

At a 5 mm shielding, the estimated total TID received by RazakSAT-1, SCD-2 and ALOS are $2.30 \times 10^2$ rads, $1.78 \times 10^3$ rads and $2.42 \times 10^4$ rads, respectively. The dose-depth curve flattens above 4 mm and 10 mm for SCD-2 and ALOS, respectively, which further implies that thicker shielding becomes less effective in reducing the dose. Thus, radiation-hard components are suggested such that a lower shielding thickness can be applied, and the mass/volume ratio can also be improved.

In terms of materials, Table 3 shows the dose comparison between silicon (Si) and GaAs, which are commonly used in spacecraft. Here, the Si material exhibited a higher dose than the GaAs material. Hence, considering the two commonly used materials in spacecraft, the GaAs material is expected to have better resistance against radiation damage than the Si material.

## 3.4 Solar-Cell Analysis

Figure 10 shows the effective dose of 1 MeV electrons as a function of the shielding thickness for GaAs solar cells in three satellite orbits over three years. Similar to other doses, the increase in the cover-glass thickness reduces the radiation damage to solar cells. The damage to the solar cells on all three satellites can be reduced by more than 50% with the use of 0.1 mm silica glass. Among the three satellites, the RazakSAT-1 solar cell received the lowest expected dose, whereas ALOS received the highest dose. However, the radiation dose reduction becomes less significant as the shield thickness increases above 0.2 mm for both RazakSAT-1 and SCD-2 and above 0.5 mm for ALOS, although their doses continue to slowly decrease.

The performance of the solar cells was based on their maximum power capability and the influence of the silica glass thickness. This characteristic was demonstrated using GaAs material as shown in Figure 11. Comparing the three satellites, the solar cells in RazakSAT-1 were expected to attain the highest power capability during the three-year mission, followed by SCD-2 and ALOS. Only a very thin silica glass is required by RazakSAT-1 to protect its solar cells, where 0.01 mm can improve the performance by 0.05% (this satellite is expected to experience the least radiation damage). Increasing the cover-glass thickness does not significantly change the outcome. For the other two satellites, the 0.01 mm silica glass can increase the performance of SCD-2 and ALOS by 0.40% and 2.19%, respectively. RazakSAT-1 can perform 0.57% and 2.62% better than SCD-2 and ALOS, respectively; thus, these latter two satellites possibly degrade faster over their mission lifetimes. In fact, despite the estimated drawback, SCD-2 and ALOS remain functional.

Figure 12 shows a comparison of the degradation of GaAs with the Si material in the RazakSAT-1 orbit. Similar to Figure 11, the power of the solar cells can be improved by increasing the cover-glass thickness, and GaAs solar cells are expected to offer a superior power performance than Si solar cells in the NEqO. Moreover, GaAs material is expected to increase the power of the solar cell by approximately 29% more than the Si material when it is shielded with 0.5 mm of silica glass. Hence, GaAs, which is used in RazakSAT-1, is a superior material because it appears to provide higher resistance to radiation damage in orbit.

## 3.5 Geomagnetic Condition Analysis

The geomagnetic conditions when communication with RazakSAT-1 was disrupted are shown in Figure 13. The three NOAA satellites had similar flux trends, and all three satellites detected sudden peak fluxes on August 4, 2010. At this point, the $K_p$ value was 4, and the $Dst$ index was -49 nT. Although the values may indicate possible degradation of the high-frequency band, the geomagnetic conditions for the entire three months were mild and insufficient to cause permanent damage to the communication devices. Moreover, the geomagnetic disturbance described by the $K_p$ and $Dst$ indices are moderately consistent with the trapped-particle fluxes of all three NOAA satellites as simplified in Table 4. Here, the correlation between trapped particles and the geomagnetic indices was 0.6-0.7, i.e., 60-70% of the data are related.

Apart from the data of the NOAA/POES satellites and geomagnetic indices, Figure 14 demonstrates the monthly fluxes of the GCR and trapped particles in the RazakSAT-1 orbit, which were estimated by SPENVIS models from 14 July 2009 to 14 July 2012. The GCR fluxes remain anti-correlated with the trapped-particle fluxes, whereas the models did not estimate any sudden anomalies over the three years. Thus, the NOAA/POES and geomagnetic source observations and the model data suggest that factors other than space radiation may be responsible for the communication failure of RazakSAT-1.

## 4. Summary and Conclusion

The radiation environment in the LEO/NEqO of the RazakSAT-1 satellite was assessed by evaluating the fluxes of three space radiation sources: GCR, SEPs and trapped particles. The fluxes of GCR, SEPs, trapped protons and trapped electrons were predicted using the CREME96, ESP-PSYCHIC, AP-8 and AE-8 models, respectively. GCR dominated the LEO/NEqO in the high-energy range of $10^3$-$10^4$ MeV, whereas no SEPs were detected because of the effective shielding effect. The maximum GCR flux in the LEO/NEqO was approximately 3,333 times less than that in the polar orbit. For trapped particles, the energies of trapped protons and electrons were below 400 MeV and 4 MeV, respectively. The fluxes of trapped protons and trapped electrons in the NEqO were 100 and 1,000 times less, respectively, than those in the polar orbit.

In terms of the radiation dose, the maximum LET values for RazakSAT-1, SCD-2 and ALOS were 1087 MeV cm$^2$ g$^{-1}$, 1162 cm$^2$ g$^{-1}$ and 21,277 cm$^2$ g$^{-1}$, respectively. RazakSAT-1 is estimated to experience 1.30-9.40 times less LET damage than SCD and 3-120 times less LET damage than ALOS. The ionization that occurred on all three satellites is mainly attributed to protons; the result is greater nuclear interactions. RazakSAT-1 has the lowest SEU rates, which are 8.01 and 5.33 times lower than those of SCD-2 and ALOS, respectively.

Meanwhile, RazakSAT-1 also had the lowest estimated TID of $2.30 \times 10^2$ rads over three years with 5 mm Al shielding; this value is approximately 7 and 105 times less than the TID of SCD-2 and ALOS respectively. However, all three satellites show that further increases in the shielding thickness do not significantly reduce the dose level.

For solar cells, 0.01 mm of silica glass can improve the performance of RazakSAT-1, SCD-2 and ALOS by 0.05%, 0.40% and 2.19%, respectively. The solar cells of RazakSAT-1 are not significantly affected in orbit: the 0.01 mm of cover glass was sufficient to produce a 0.57% and 2.62% better performance than SCD-2 and ALOS, respectively. Furthermore, the GaAs material that was used in RazakSAT-1 also provides 29% better radiation resistance than Si material with 0.5 mm silica glass.

The environmental data from NOAA-15, 16 and 17 and the geomagnetic indices indicate mild conditions during the three-month disruption period, except for a sudden peak on 4 August 2010, whose corresponding $K_p$ index was 4. However, this event was insufficient to cause long-term damage to the communication frequency band of RazakSAT-1. The estimated monthly fluxes of GCR and trapped particles from the SPENVIS models did not experience any anomalies during RazakSAT-1 three-year mission. Thus, factors other than space radiation may be responsible for the failure of RazakSAT-1. These factors include technical failure, erosion by atomic oxygen, and micrometeoroids impacts. However, this hypothesis requires further study.

**Conflict of Interests**

The authors declare that there is no conflict of interests regarding the publication of this paper.

**Acknowledgments**

This research is supported by the Ministry of Education, Malaysia (MoE) under grant FRGS/2/2013/SG02/UKM/02/3. We would like to thank the SPENVIS team, NOAA and ISGI for providing the software, data on trapped particles and data on geomagnetic indices, respectively.

**References**

[1] A. Ahmad, "Classification simulation of RazakSAT satellite", *Procedia Engineering for Malaysian Technical Universities Conference on Engineering and Technology 2012 (MUCET 2012)*, vol. 53, pp. 472-482, 2013.

[2] M. Hashim, M. S. El-Mahallawy, M. N. M. Reba, A. A. Abas, S. Ahmad, X. Q. Yap, M. Marghany, and A. S. Arshad, "Geometric and radiometric evaluation of RazakSAT medium-sized aperture camera data", *International Journal of Remote Sensing*, vol. 34, no. 11, pp. 3947-3967, 2013.

[3] H. J. Chin, B. J. Kim, H. S. Cheng, E. E. Kim, W. K. Park, S. D. Park, and A. S. Ashard, "RazakSAT- a high performance satellite waiting for its mission in space", *20$^{th}$ Annual AIAA/USU Conference on Small Satellites*, Article ID SSC06-VI-6, 2006.

[4] S. W. Samwel and A. A. Hady "Space radiation environment forecast for EGYPTSAT-2 satellite", *Space Weather*, vol. 7, Article ID S12004, 2009.

[5] D. Heynderickx, B. Quaghebeur, J. Wera, E. J. Daly, and H. D. R Evans, "New radiation environment and effects models in the European Space Agency's Space

Environment Information System (SPENVIS)", *Space Weather*, vol. 2, no. 10, Article ID 10.1029/2004SW000073, 2004.

[6] M. A. Xapsos, C. Stauffer, T. Jordan, J. L. Barth, and R. A. Mewaldt, "Model for cumulative solar heavy ion energy and linear energy transfer spectra", *IEEE Transactions on Nuclear Science*, vol. 54, no. 6, pp. 1985-1989, 2007.

[7] D. M. Sawyer and J. I. Vette, "AP-8 Trapped proton environment for solar maximum and solar minimum", NSSDC Report 76-06, 1976.

[8] J. I. Vette, "The AE-8 trapped electron model environment", NSSDC Report 91-24, 1991.

[9] J. H. Adams, "Cosmic ray effects on microelectronics, Part IV", NRL Memorandum Reports 5901, 1986.

[10] J. H. Adams, J. Bellingham, and P. E. Graney, "A comprehensive table of ion stopping powers and ranges", NRL Memorandum Report, 1987.

[11] S. M. Seltzer, "SHIELDOSE, a computer code for space-shielding radiation dose calculations, national bureau of standards", NBS Technical Note 1116, U.S. Government Printing Office, Washington, D.C., 1980.

[12] S. M. Seltzer, "Updated calculations for routine space-shielding radiation dose estimates: SHIELDOSE-2", NIST Publication NISTIR, no. 5477, 1994.

[13] W. H. Barkas, and M. J. Berger, "Tables of energy losses and ranges of heavy charged particles", NASA Technical Report SP-3013, 1964.

[14] R. G. Alsmiller, J. Barish, and W. W. Scott, *Nuclear Science and Engineering*, vol. 35, pp. 303-318, 1969.

[15] H. Y. Tada, J. R. Carter, Jr. B. E. Anspaugh and R. G. Downing, "Solar cell radiation handbook", Third Edition, JPL Publication, pp. 82-69, 1982.

[16] S. R. Messenger, R. J. Walters, J. H. Warner, H. Evans, S. J. Taylor, C. Baur, and D. Heynderickx, "Status of implementation of displacement damage dose method for space solar cell degradation analyses", *8th European Space Power Conference*, 2008.

[17] R. A. Nymmik, M. I. Panasyuk, and A. A. Suslov, "Galactic cosmic ray flux simulation and prediction", *Nuclear Tracks and Radiation Measurements*, vol. 20, pp. 427-429, 1996.

[18] E. R. Benton and E.V. Benton, "Space radiation dosimetry in low-earth orbit and beyond", *Nuclear Instruments and Methods in Physics Research*, vol. 184 (B), pp. 255-294, 2001.

[19] H. Murtaza, "Prediction of the space radiation environment of PakSat, a geostationary communication satellite", *Journal of Space Technology*, vol. 1, no. 1, pp. 73-77, 2011.

**Figure Captions**

Figure 1. Block diagram of the procedures used in SPENVIS to predict the space radiation environment for the three-year mission (2009-2012)

Figure 2. GCR spectra generated from four SPENVIS models for **a)** a NEqO, which is represented by RazakSAT-1 and **b)** a polar orbit, which is represented by ALOS

Figure 3. Comparison of GCR fluxes for satellites with different inclinations

Figure 4. Solar particles estimated by SPENVIS in the polar ALOS orbit for the three-year mission

Figure 5. **a)** Energy spectra of the trapped protons and **b)** trapped electrons for the three-year RazakSAT-1 mission

Figure 6. **a)** Comparison of the trapped protons and **b)** trapped electrons among three orbital datasets for the three-year mission. RazakSAT-1 is at 9°, SCD-2 is at 25°, and ALOS is at 98.16°

**Table Captions**

Table 1. Orbital data of the three satellites that represented the polar orbit and NEqO for the three-year mission from 14 July 2009 to 14 July 2012

Table 2. SEU rates for 5 mm Al shielding through a GaAs device for three satellites over three years

Table 3. Absorbed dose for two materials shielded with selected thickness values based on RazakSAT-1 orbital data for a total mission period of three years

Table 4. Correlation coefficient between the trapped particles with the $K_p$ and $Dst$ indices for three NOAA satellites from June to August 2010

Figure 1

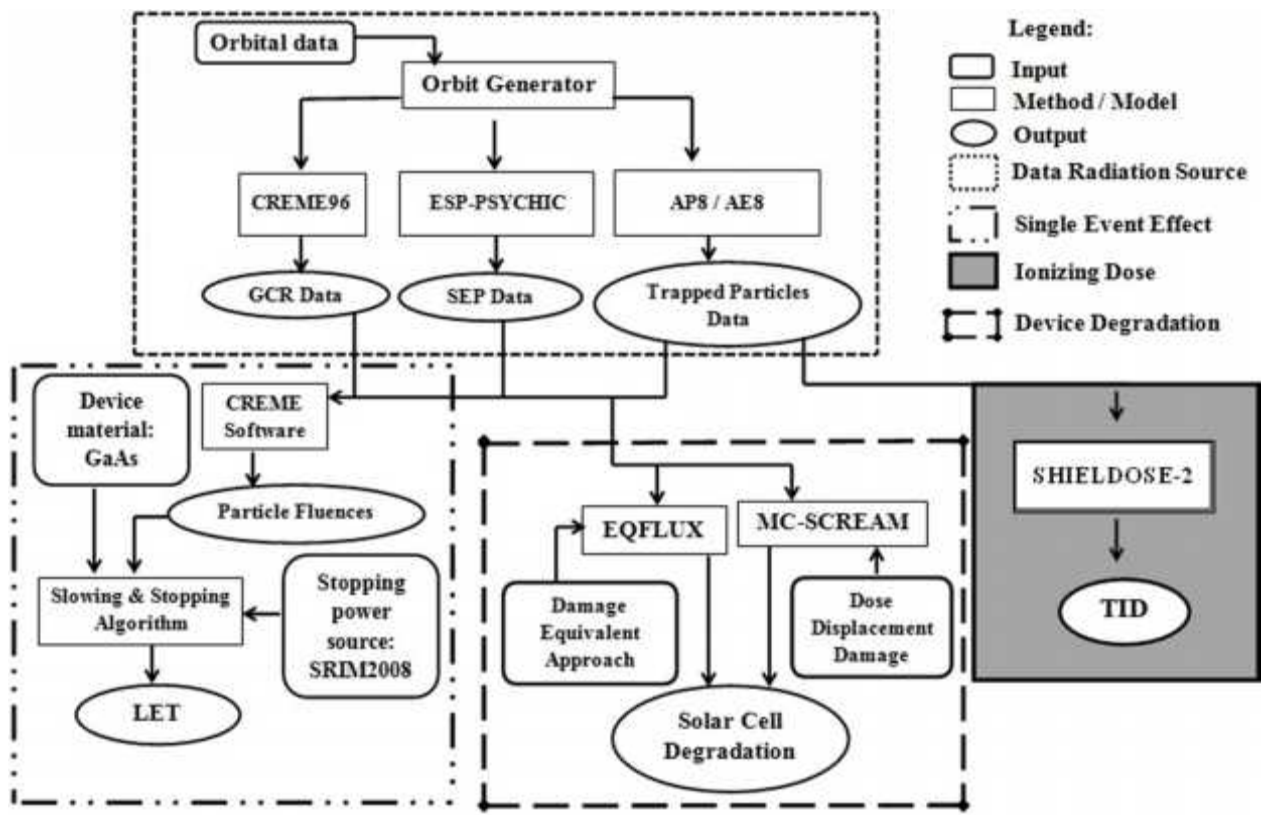

Figure 2

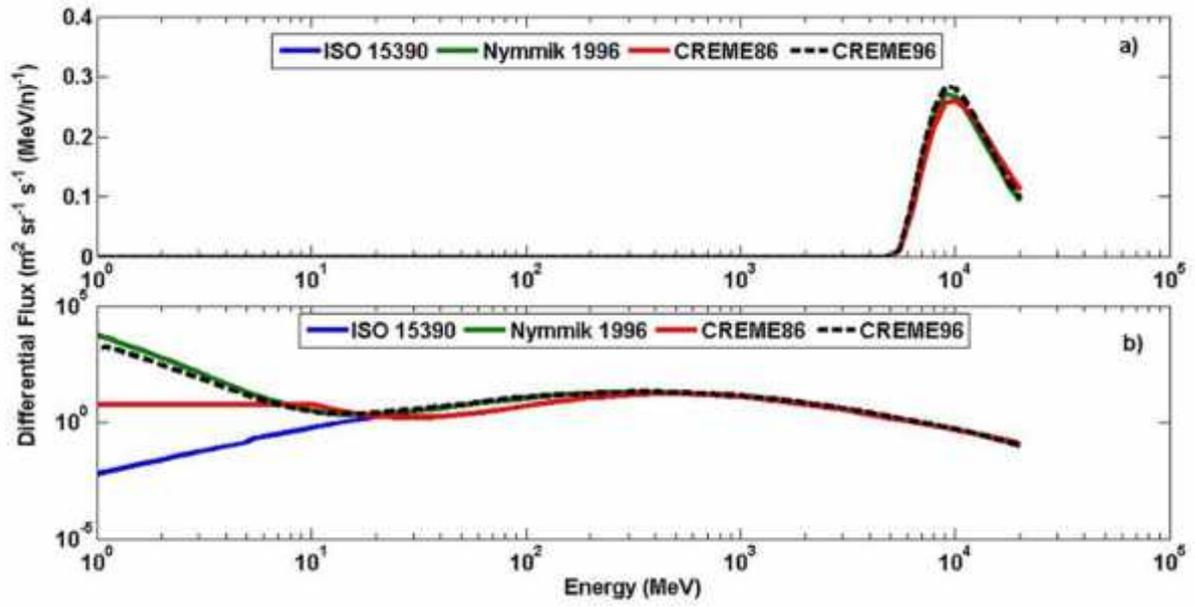

Figure 3

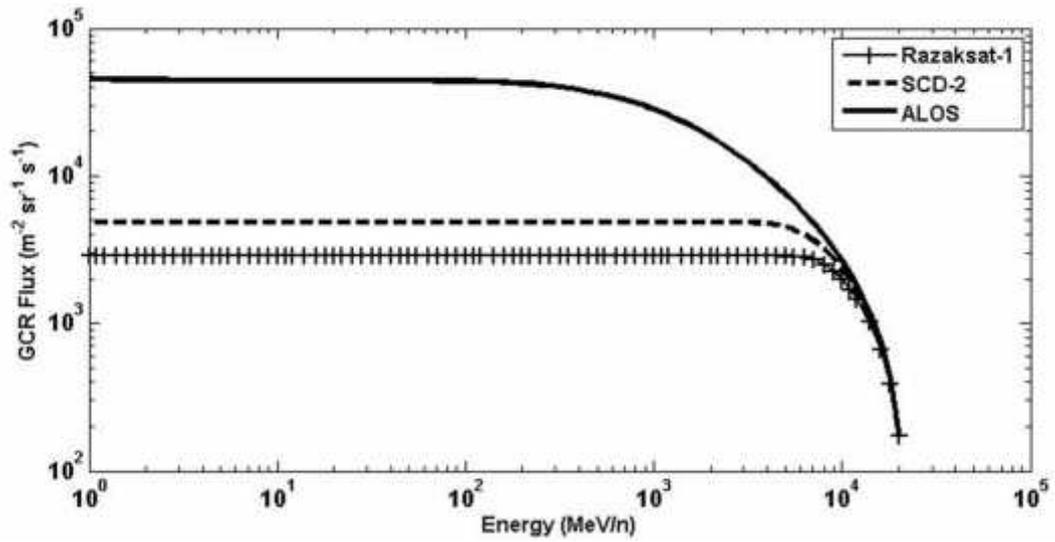

Figure 4

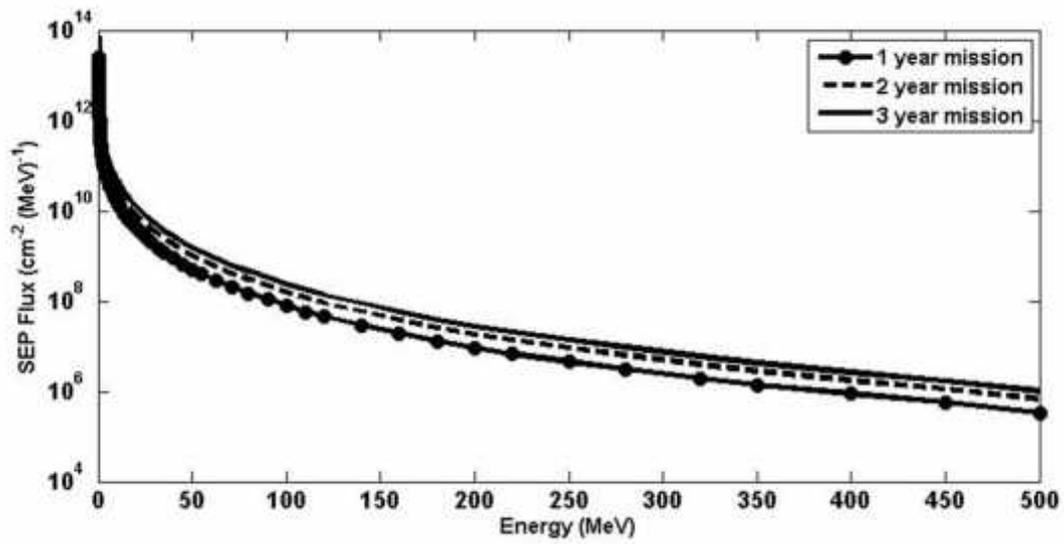

Figure 5

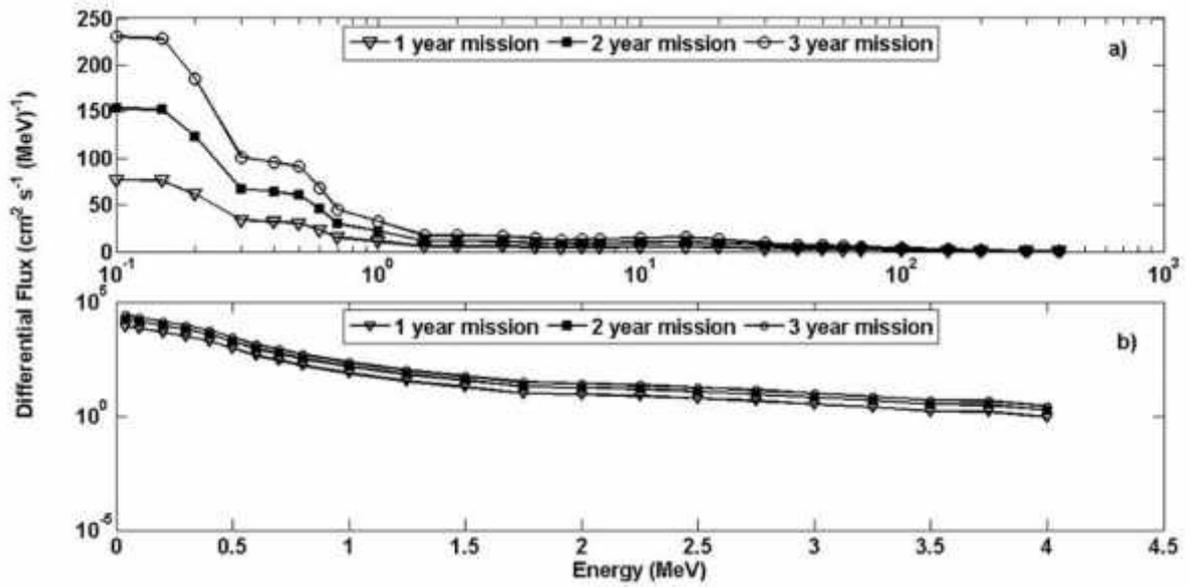

Figure 6

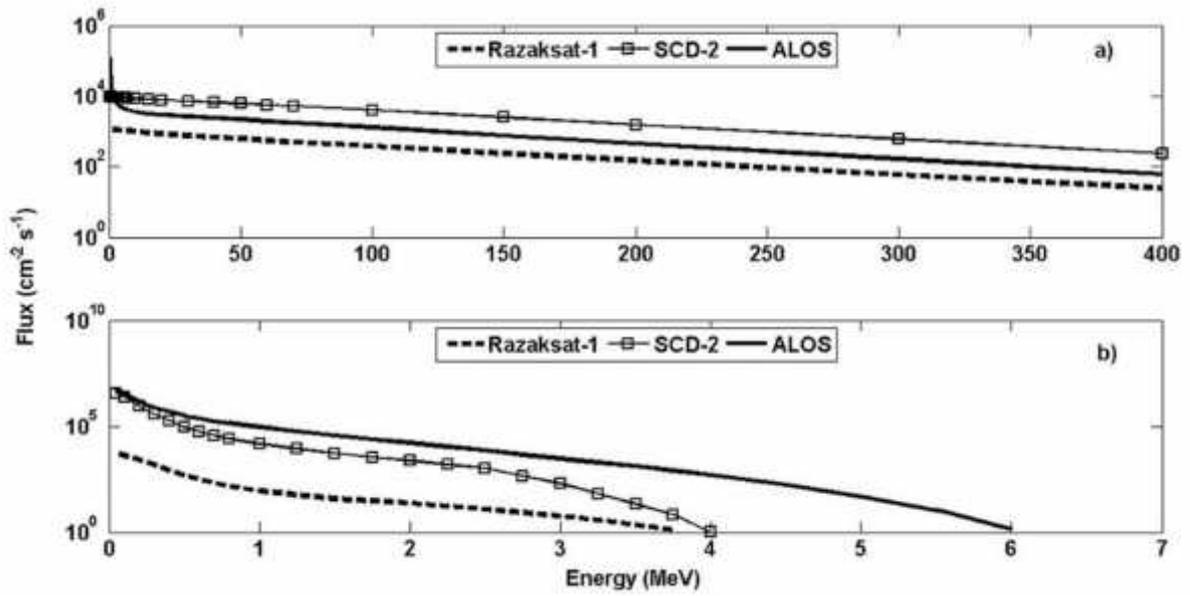

Figure 7

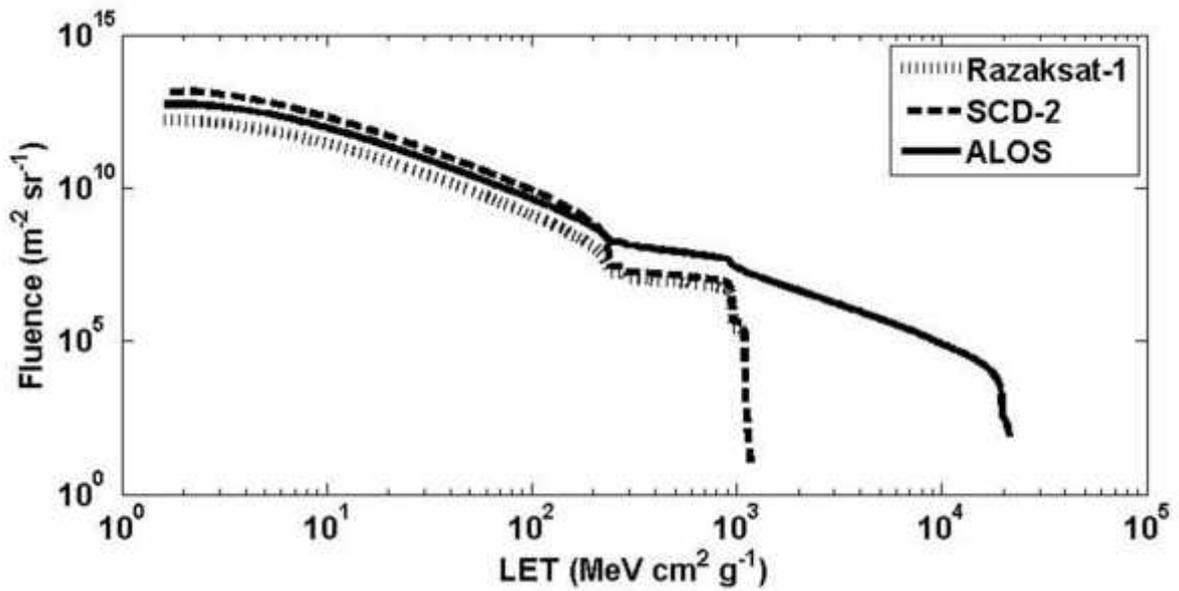

Figure 8

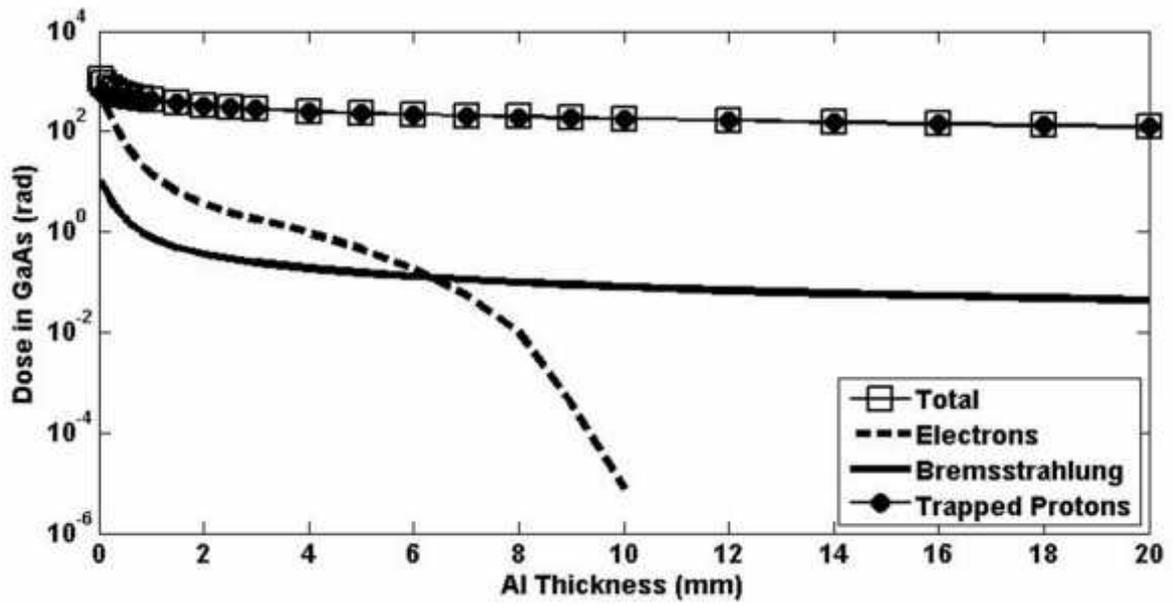

Figure 9

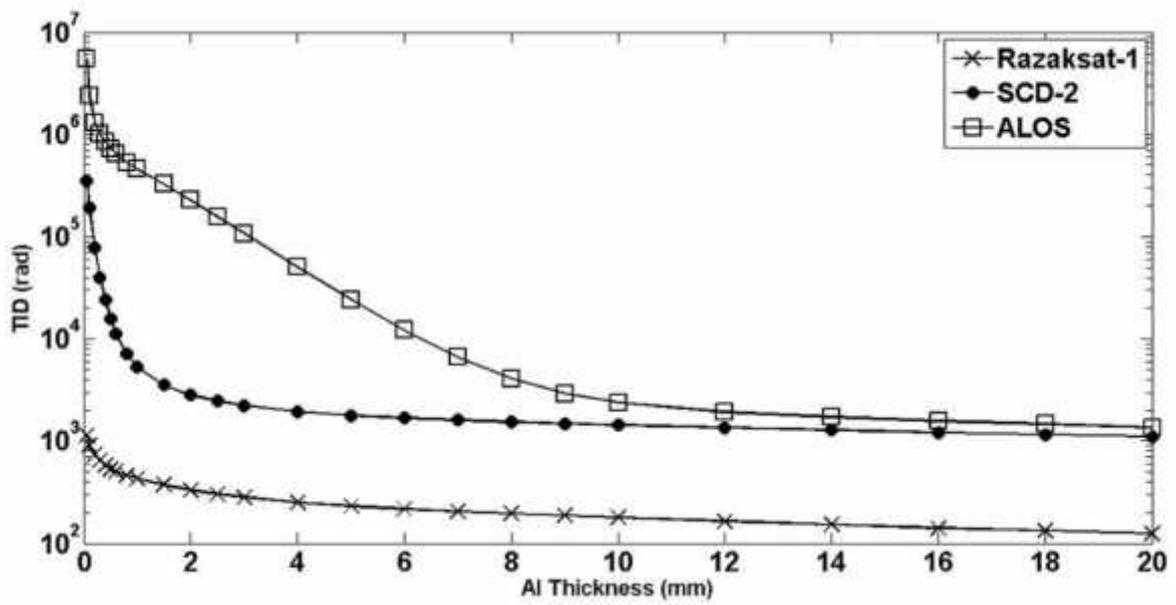

Figure 10

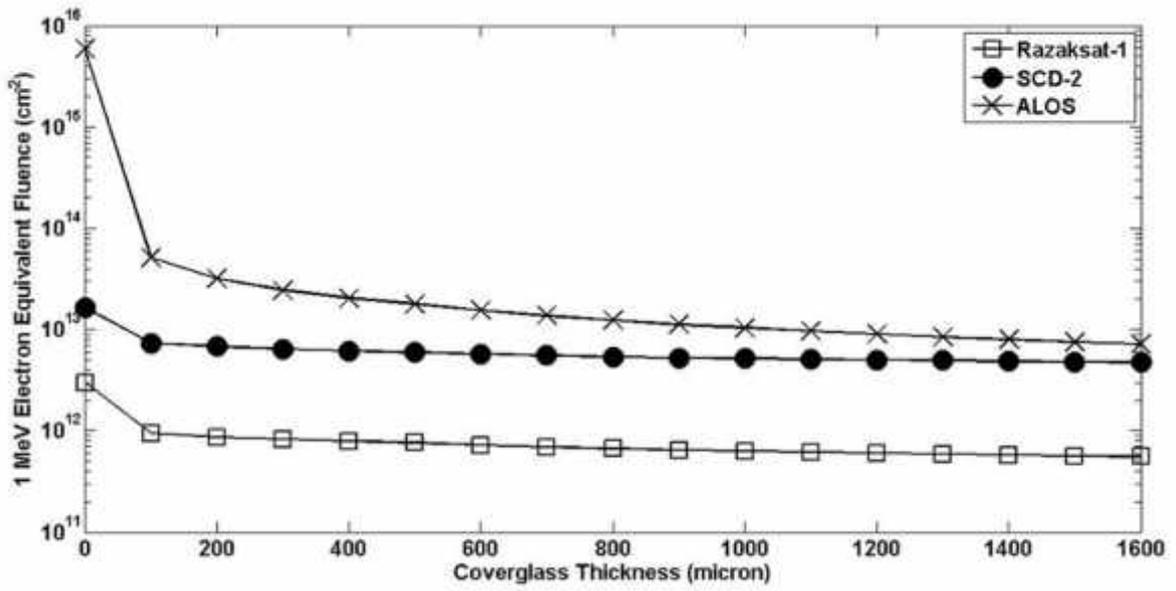

Figure 11

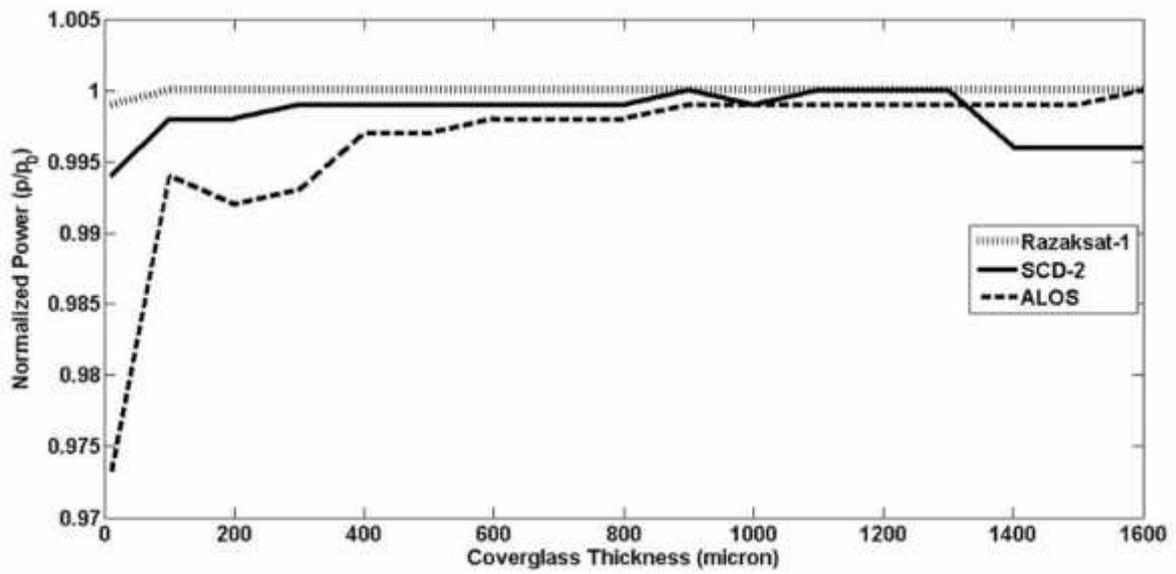

Figure 12

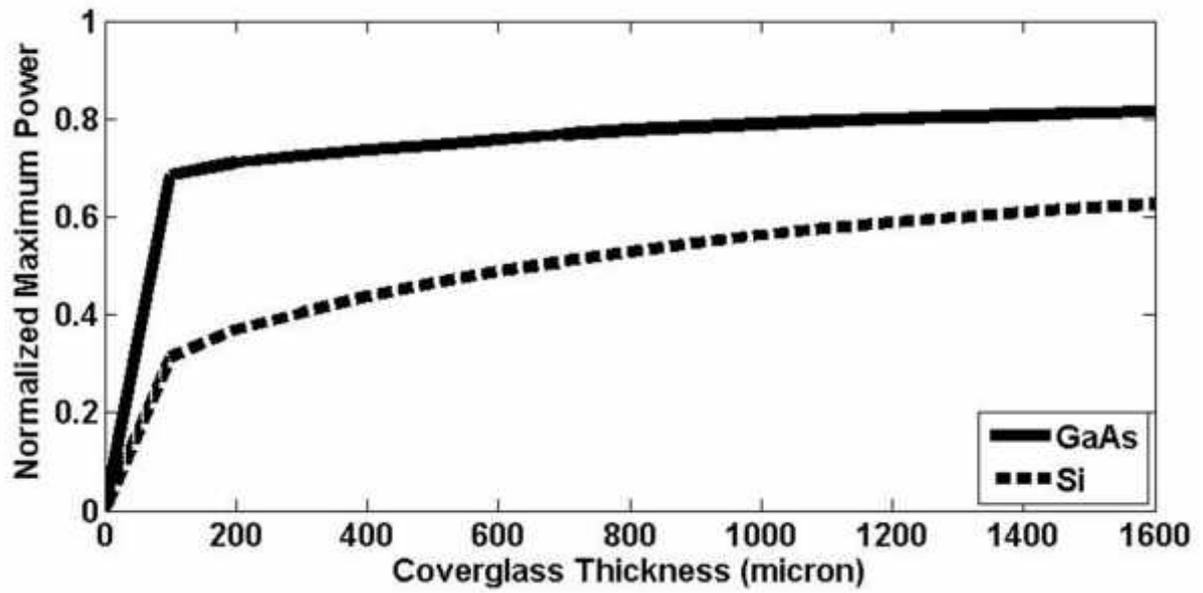

Figure 13

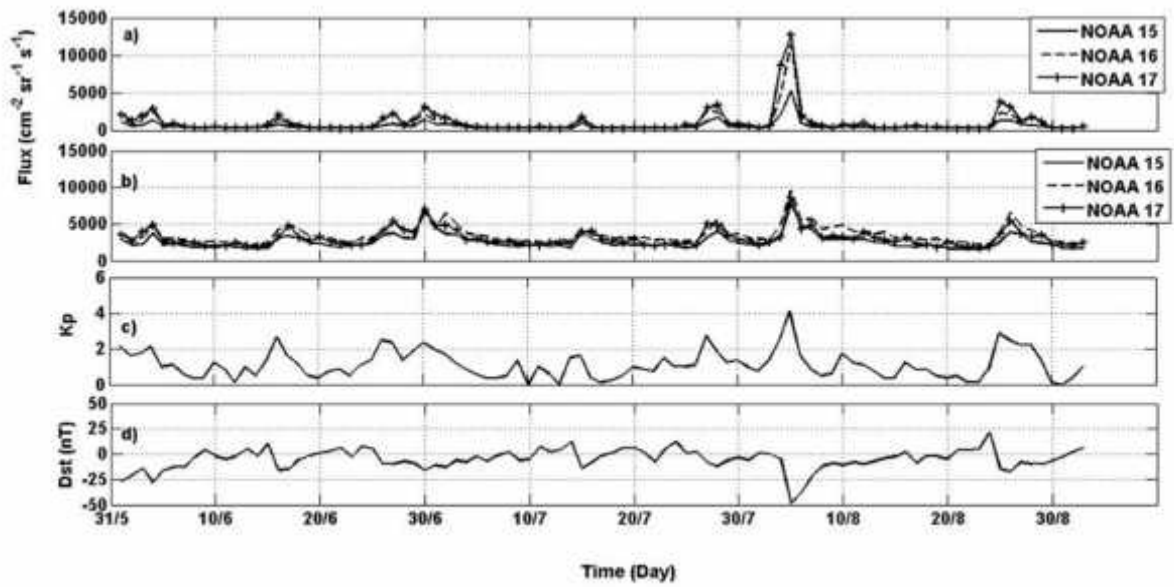

Figure 14

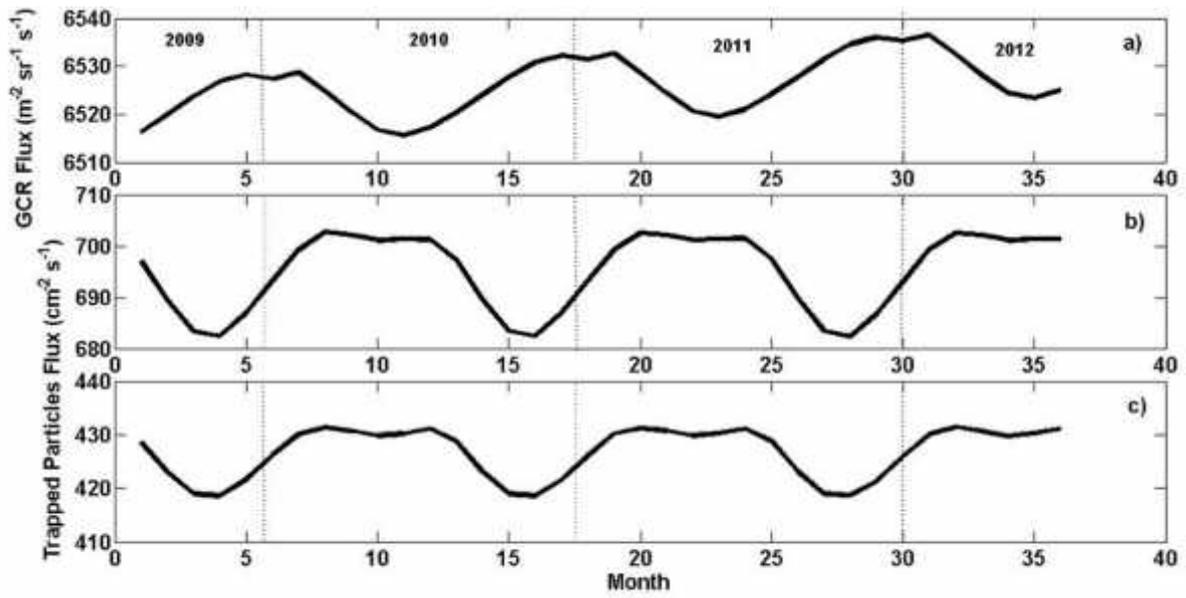

Table 1. Orbital data of three satellites used to represent polar and NEqO orbits for three years mission period from 14 July 2009 to 14 July 2012

| Satellite | NEqO Orbit Razaksat-1 | NEqO Orbit SCD-2 | Polar Orbit ALOS |
| --- | --- | --- | --- |
| Perigee (km) | 662 | 743 | 691 |
| Apogee (km) | 687 | 768 | 693 |
| Inclination (degree) | 9° | 25° | 98.16° |

Table 2. SEU rates at 5 mm Al shielding through a GaAs device for three different satellites for a total mission of 3 years

| Effect/Satellite | Razaksat-1 | SCD-2 | ALOS |
|---|---|---|---|
| Direct Ionization (bit$^{-1}$) | $2.13 \times 10^{-2}$ | $3.52 \times 10^{-2}$ | $1.89 \times 10^{-1}$ |
| Proton-Induced Ionization (bit$^{-1}$) | $6.48 \times 10^{-2}$ | $6.54 \times 10^{-1}$ | $2.70 \times 10^{-1}$ |
| Total Ionization (bit$^{-1}$) | $8.61 \times 10^{-2}$ | $6.89 \times 10^{-1}$ | $4.59 \times 10^{-1}$ |

Table 3. Absorbed dose for two different materials shielded by selected thickness following Razaksat-1 orbital data for a total mission of 3 years

| Al Thickness (mm) | Total TID (rad) | |
|---|---|---|
| | GaAs | Si |
| 1 | $4.29 \times 10^2$ | $5.84 \times 10^2$ |
| 2 | $3.32 \times 10^2$ | $4.47 \times 10^2$ |
| 3 | $2.82 \times 10^2$ | $3.77 \times 10^2$ |
| 4 | $2.52 \: 10^2$ | $3.35 \times 10^2$ |
| 5 | $2.30 \times 10^2$ | $3.05 \times 10^2$ |
| 6 | $2.16 \times 10^2$ | $2.86 \times 10^2$ |

Table 4. Correlation coefficient between trapped particles with $K_p$ and Dst indices for three NOAA satellites for a period of 3 months (June, July and August 2010)

| Particle | Satellite | Correlation Coefficient | |
|---|---|---|---|
| | | $K_p$ index | Dst index |
| Trapped Protons | NOAA-15 | 0.7275 | -0.6422 |
| | NOAA-16 | 0.7335 | -0.6350 |
| | NOAA-17 | 0.7480 | -0.5968 |
| Trapped Electrons | NOAA-15 | 0.6654 | -0.7114 |
| | NOAA-16 | 0.6844 | -0.7346 |
| | NOAA-17 | 0.7211 | -0.7125 |